\documentstyle[12pt]{article}

\def\nqq{\hspace{-2em}}

\def\barr{\left(\begin{array}}
\def\earr{\end{array}\right)}
\def\beq#1{\begin{equation}\label{#1}}
\def\eeq{\end{equation}}
\def\ber#1{\begin{eqnarray}\label{#1} \nqq}
\def\eer{\end{eqnarray}}
\def\eern{\nonumber \end{eqnarray}}
\def\nn{\nonumber\\ \nqq}
\def\mm{\\ \nqq}

\catcode`\@=11 \@addtoreset{equation}{section}\catcode`\@=12

\newcommand{\R}{\mbox{\bf R}}
\newcommand{\C}{\mbox{\bf C}}
\newcommand{\Z}{\mbox{\bf Z}}
\newcommand{\N}{\mbox{\bf N}}
\newcommand{\SL}{\mathop{\rm sl}\nolimits}
\newcommand{\SO}{\mathop{\rm so}\nolimits}
\newcommand{\const}{\mathop{\rm const}\nolimits}
\newcommand{\rank}{\mathop{\rm rank}\nolimits}
\newcommand{\diag}{\mathop{\rm diag}\nolimits}
\newcommand{\sign}{\mathop{\rm sign}\nolimits}

\newcommand{\eps}{\varepsilon}
\newcommand{\tri}{\triangle}

\newcommand{\e}[1]{\mathop{\rm e}\nolimits^{#1}}

\newcommand{\p}{\partial}

\newcommand{\fnm}{\footnotemark}
\newcommand{\fnt}{\footnotetext}

\begin{document}

\begin{center}
\large\bf
MAJUMDAR-PAPAPETROU TYPE SOLUTIONS IN \\
SIGMA-MODEL AND INTERSECTING P-BRANES \\[15pt]

\normalsize\bf V.D. Ivashchuk\fnm[1]\fnt[1]{ivas@rgs.phys.msu.su}
and
V.N. Melnikov\fnm[2]\fnt[2]{rgs@com2com.ru} \\[10pt]

\it
Center for Gravitation and Fundamental Metrology,
VNIIMS \\ 3-1, M.Ulyanovoy Str., Moscow 117313, Russia
\end{center}

\vspace{15pt}


\begin{abstract}
The block-orthogonal generalization of the Majumdar-Papapetrou
type solutions for the $\sigma$-model studied earlier
are obtained and corresponding solutions with $p$-branes are considered.
The existence of solutions and the number of independent harmonic
functions is defined by the matrix of scalar products of vectors
$U^s$, governing the $\sigma$-model target space metric. For orthogonal
$U^s$, when target space is a symmetric homogeneous space, the solutions
reduce to the previous ones. Two special
classes of obtained solutions
with $U^s$ related to finite dimensional Lie algebras and hyperbolic
(Kac-Moody) algebras are singled out and investigated. The affine
Cartan matrices do not arise in the scheme under consideration.
Some examples of solutions and intersection rules for $D=11$
supergravity, related $D=12$ theory and extending them $B_D$-models
are considered. For special multicenter solutions criterions
for the existence of horizon and curvature singularity are found.
\end{abstract}

\pagebreak

\section{Introduction}

In this paper we consider a family of Majumdar-Papapetrou
(MP) type   solutions \cite{MP} for gravitating sigma-model (for review
see, for example, \cite{Cher}) with the action
\ber{1.1a}
S_\sigma=\int
d^{d_0}x\sqrt{|g^0|}\biggl\{R[g^0]-\hat G_{AB}
g^{0\mu\nu}\p_\mu\sigma^A\p_\nu\sigma^B   - 2V(\sigma)
-\sum_{s\in S}\eps_s\e{-2U_A^s\sigma^A}
g^{0\mu\nu} \p_\mu\Phi^s\p_\nu\Phi^s\biggr\},
\eer
where $g^0=g^0_{\mu\nu}(x)dx^\mu\otimes dx^\nu$ is a metric on
$d_0$-dimensional manifold $M_0$,
$R[g^0]$ is the scalar curvature of $g^0$,
$\sigma=(\sigma^A)\in\R^N$ and
$\Phi=(\Phi^s,s\in S)$ is a set of  scalar fields ($S\ne\emptyset$),
$V=V(\sigma)$ is a potential, $(\hat G_{AB})$ is a symmetric
non-degenerate matrix, $U^s=(U_A^s)\in\R^N$ are vectors
and $\eps_s=\pm1$, $s\in S$.
The sigma-model (\ref{1.1a}) arises in multidimensional gravitational
models with scalar fields and fields of forms, when solutions with
intersecting $p$-branes are considered \cite{IM4}-\cite{IMR}
(a pure gravitational sector of the sigma-model
was considered in \cite{Ber,RZ,IM0}).
For $p$-brane applications (see, for example, \cite{IM4}-\cite{IMJ})
$g^0$ is Euclidean,
$(\hat G_{AB})$ is positively defined and
$\eps_s= -1$, if pseudo-Euclidean (electric and magnetic) $p$-branes
in a pseudo-Euclidean space-time are considered.
The sigma-model (\ref{1.1a}) may be also considered
for the pseudo-Euclidean metric $g^0$ of signature
$(-,+, \ldots, +)$. In this case
for a positively defined matrix $(\hat G_{AB})$ and
$\eps_s= 1$ we get a non-negative kinetic energy terms.

The target space of the model is $(\R^K,{\cal G})$, where
\ber{1.1}
{\cal G}=\hat G_{AB}d\sigma^A\otimes d\sigma^B+
\sum_{s\in S}\eps_s\e{-2U_A^s\sigma^A}d\Phi^s\otimes d\Phi^s,
\eer
It was proved in \cite{Cosm} that the target space
${\cal T}=(\R^K,{\cal G})$ is a homogeneous (coset) space $G/H$
($G$ is the isometry group of ${\cal T}$,
$H$ is the isotropy subgroup of $G$). ${\cal T}$ is
symmetric (i.e. the Riemann tensor is covariantly constant:
$\nabla_MR_{M_1M_2M_3M_4}[{\cal G}]=0$) if and only if
\ber{1.2}
(U^{s_1}-U^{s_2})(U^{s_1},U^{s_2})=0
\eer
for all $s_1,s_2\in S$, i.e. when any two vectors $U^{s_1}$ and $U^{s_2}$,
$s_1\ne s_2$, are either coinciding $U^{s_1}=U^{s_2}$ or orthogonal
$(U^{s_1},U^{s_2})=0$, where scalar product
$(\cdot,\cdot)$ is defined as follows
\ber{1.2a}
(U,U')=\hat G^{AB}U_AU'_B,
\eer
for $U,U'\in\R^N$, where $(\hat G^{AB})=(\hat G_{AB})^{-1}$.

In our previous papers \cite{IM4}-\cite{IMR}
we considered the orthogonal case
\ber{1.3}
(U^{s_1},U^{s_2})=0,
\eer
$s_1\ne s_2$, and $(U^{s_1},U^{s_1})\ne0$,
$s_1,s_2\in S$, and for the zero-potential:
$V=0$, obtained a family of exact solutions in the
 sigma-model (\ref{1.1a})
governed by $k$ harmonic functions, where $k$ is a number of $s$, satisfying
$\eps_s(U^s,U^s)<0$. Using these sigma-model solutions we also
found a family of
solutions in multidimensional gravity with $p$-branes.
For $p$-brane applications the orthogonality condition  (\ref{1.3})
is equivalent to (orthogonal) intersection rules
\cite{IM4,IM,IMC,IMR,AR,AEH} also known as no-force condition.

In \cite{IMC} these solutions (sigma-model and $p$-brane ones)
were also generalized  to the case $V\ne0$, where
$V=\sum_{a=1}^m A_a\e{u_A^a\sigma^A}$ and $(u^a,U^s)=0$, $a=1,\dots,m$,
$s\in S$.

Here we generalize the orthogonal
$\sigma$-model solutions from \cite{IM4}-\cite{IMR} to
a more general block-orthogonal
case:
\ber{2.3}
S=S_1 \cup\dots\cup S_k, \qquad  S_i \cap S_j = \emptyset, \quad i \neq j
\eer
$S_i \ne \emptyset$, i.e. the set $S$ is a union of $k$ non-intersecting
(non-empty) subsets $S_1,\dots,S_k$,
and
\ber{2.4}
(U^s,U^{s'})=0
\eer
for all $s\in S_i$, $s'\in S_j$, $i\ne j$; $i,j=1,\dots,k$.
According to (\ref{2.4}) the set of
vectors $(U^s,s\in S)$ has a block-orthogonal structure with respect to
the scalar product (\ref{1.2a}): it is splitted into $k$ mutually
orthogonal blocks $(U^s,s\in S_i)$, $i=1,\dots,k$.

In this case solutions do exist, when the matrix of scalar products
$B = ((U^{s_1},U^{s_2}))$ and $\eps_s$ satisfy the relation
\ber{2.10}
\sum_{s'\in S}(U^s,U^{s'})\eps_{s'}\nu_{s'}^2=-1
\eer
for some set of real $\nu_s$, $s \in S$.
The number of independent harmonic functions is defined by the number of
blocks in the matrix $B$.

The paper is organized as follows. In Sect. 2 the sigma-model
is considered and in block-orthogonal case exact solutions
governed by a set of  harmonic functions are obtained. In  Sect. 3 certain
examples of solutions related to Lie algebras (e.g. finite dimensional and
hyperbolic) are considered and restrictions
on signature parameters $\eps_s$ for different Lie algebras
 are derived. It is shown that
affine Kac-Moody algebras do not appear in this scheme.
Sect. 4 is devoted  to application of the sigma-model solutions
to multidimensional models with intersecting p-branes.
The block-orthogonal generalization of (orthogonal) p-brane
MP type solutions is presented in subsect. 4.2
(We note, that recently block-orthogonal black hole and wormhole
solutions with p-branes were considered in \cite{Br}.
These solutions generalize orthogonal black hole solutions
\cite{CT,OS,AIV,O,BIM,IMJ}.)
In subsect. 4.3  the behavior of the Riemann tensor squared  (Kretschmann
scalar) for multicenter  solutions is investigated and criteria for the
existence of horizon and finiteness of the Kretschmann scalar are
established.  In Sect. 5  intersection rules and some examples are
considered (e.g. for $B_D$-models that may be relevant for future
generalizations of $M$- and $F$-theories \cite{M-th,F-th} to dimensions $D >
12$.) Here a dyon solution for $D=11$ supergravity with electric 2-brane and
magnetic 5-brane configuration is considered.  This solution has $A_2 =
sl(3,\C)$ intersection rule:  $3 \cap 6 = 1$ instead of the orthogonal one
$3 \cap 6 = 2$.  An analogous dyon solution for $D=10$  IIA supergravity was
recently considered in \cite{GrI}.

We note that the  intersecting $p$-brane solutions
(see \cite{IM4,IM,IMC,IMR,AV,AR,AEH,St} and references therein)
with ``orthogonal'' intersection rules correspond to the
Lie algebras $A_1 \oplus \ldots \oplus A_1$,
where $A_1 =sl(2,\C)$. (The MP solution in this
classification corresponds to the algebra $A_1$.)
In supergravitational  models these solutions correspond to the
so-called BPS saturated states preserving fractional supersymmetry
\cite{DS}-\cite{Ga}.

\section{$\sigma$-model solutions}

Here we considers a family of solutions to equations of motion
of the sigma-model (\ref{1.1a}) in the block-orthogonal
case (\ref{2.3}), (\ref{2.4}).

Equations of motion corresponding to (\ref{1.1a}) have the following
form
\ber{2.5}
R_{\mu\nu}[g^0]=\hat G_{AB}\p_\mu\sigma^A\p_\nu\sigma^B+
\frac{2V}{d_0-2}g_{\mu\nu}^0+\sum_{s\in S}\eps_s
\e{-2U_A^s\sigma^A}\p_\mu\Phi^s\p_\nu\Phi^s, \mm
\label{2.6}
\hat G_{AB}\tri[g^0]\sigma^B-\frac{\p V}{\p\sigma^A}+
\sum_{s\in S}\eps_sU_A^s\e{-2U_C^s\sigma^C}
g^{0\mu\nu} \p_\mu\Phi^s\p_\nu\Phi^s =0, \mm
\label{2.7}
\p_\mu\left(\sqrt{|g^0|}g^{0\mu\nu}\e{-2U_A^s\sigma^A}
\p_\nu\Phi^s\right)=0,
\eer
$s\in S$. Here  $\tri[g^0]$ is the Laplace-Beltrami operator
corresponding to $g^0$.

{\bf Proposition 1.} Let $(M_0,g^0)$ be Ricci-flat
\ber{2.8}
R_{\mu\nu}[g^0]=0.
\eer
Then the field configuration
\ber{2.9}
g^0, \qquad \sigma^A=\sum_{s\in S}\eps_sU^{sA}\nu_s^2\ln H_s, \qquad
\Phi^s=\frac{\nu_s}{H_s},
\eer
$s\in S$, satisfies the field equations (\ref{2.5})--(\ref{2.7}) with
$V=0$ if real numbers $\nu_s$ obey the relations
$$
\sum_{s'\in S}(U^s,U^{s'})\eps_{s'}\nu_{s'}^2=-1
$$
$s\in S$, functions $H_s >0$ are harmonic, i.e.
\ber{2.11}
\tri[g^0]H_s=0,
\eer
$s\in S$, and $H_s$ are coinciding inside blocks:
\ber{2.12}
H_s=H_{s'}
\eer
for $s,s'\in S_i$, $i=1,\dots,k$.

The Proposition 1 can be readily verified by a straightforward
substitution of (\ref{2.8})--(\ref{2.12}) into equations of motion
(\ref{2.5})--(\ref{2.7}). In special (orthogonal) case, when any
block contains only one vector (i.e. all $|S_i|=1$) the Proposition 1
coincides with that of \cite{IMC}. In general case vectors inside each
block $S_i$ are not orthogonal. The solution under consideration
depends on $k$ independent harmonic functions. For a given set of
vectors $(U^s,s\in S)$ the maximal number $k$ arises for irreducible
block-orthogonal decomposition (\ref{2.3}), (\ref{2.4}), when any block
$(U^s,s\in S_i)$ can not be splitted into two mutually-orthogonal
subblocks.

We note that due to (\ref{2.4}) the relation (\ref{2.10}) may
be rewritten as
\ber{2.13}
\sum_{s'\in S_i}(U^s,U^{s'})\eps_{s'}\nu_{s'}^2=-1,
\eer
$s\in S_i$, $i=1,\dots,k$. Hence, parameters $(\nu_s,s\in S_i)$
depend upon vectors $(U^s,s\in S_i)$, $i=1,\dots,k$.

\section{Parameters of solutions related to Lie algebras}

Here we put
\ber{3.1}
(U^s,U^s)\ne0
\eer
for all $s\in S$ and introduce quasi-Cartan matrix $A=(A^{ss'})$
\ber{3.2}
A^{ss'} \equiv \frac{2(U^s,U^{s'})}{(U^{s'},U^{s'})},
\eer
$s,s'\in S$,
which coincides with the Cartan matrix,
when $U^s$, $s \in S$, are  simple roots  of
some Lie algebra  and $(.,.)$ is standard
bilinear form on the root space.
>From (\ref{2.4}) we get a block-orthogonal
structure of $A$:
\ber{3.3}
A=\barr{ccc}
A_{(1)} \dots 0\\
\vdots \ddots  \vdots\\
0 \dots  A_{(k)}
\earr,
\eer
where $A_{(i)}=(A^{ss'},s,s'\in S_i)$, $i=1,\dots,k$. Here we tacitly
assume that the set $S$ is ordered, $S_1<\dots<S_k$ and the order in $S_i$
is inherited by the order in $S$.

For $\det A_{(i)}\ne0$ relation (\ref{2.13}) may be rewritten in the
equivalent form
\ber{3.4}
\eps_s\nu_s^2(U^s,U^s)=-2\sum_{s'\in S_i}A_{ss'}^{(i)},
\eer
$s\in S_i$, where $(A_{ss'}^{(i)})=A_{(i)}^{-1}$. Thus, eq. (\ref{2.13})
may be resolved in terms of $\nu_s$ for certain $\eps_s=\pm1$, $s\in S_i$.
For $\det A_{(i)}=0$ there exist situations when eq. (\ref{2.13}) has no
solutions even for complex $\nu_s$. Indeed, let us suppose
that there exists a vector
$a=(a_s,s\in S_i)$ satisfying the relations
\ber{3.5}
\sum_{s\in S_i}a_s A_{(i)}^{ss'}=0, \quad
\sum_{s\in S_i}a_s \ne 0,
\eer
$s'\in S_i$ (eqs. (\ref{3.5}) imply $\det A_{(i)}=0$ and, hence,
$\det A=0$). From (\ref{2.13}) and the first relation in (\ref{3.5})
we get $\sum_{s\in S_i} a_s=0$, that contradicts the second relation
in (\ref{3.5}).

In what follows we consider the block-orthogonal decomposition to be
irreducible, i.e. for any $i$ the block $(U^s,s\in S_i)$ can not be
splitted into two mutually orthogonal subblocks. In this case any matrix
$A_{(i)}$ is indecomposable (or irreducible) in the sense that there is
no renumbering of vectors which would bring $A_{(i)}$ to the block
diagonal form $A_i=\diag(A'_{(i)},A''_{(i)})$.

Let $A$ be a generalized Cartan matrix \cite{Kac,FS}. In this case
\ber{3.6}
A^{ss'}\in -\Z_+ \equiv\{0,-1,-2,\dots\}
\eer
for $s \ne s'$ and $A$ generates generalized
symmetrizable Kac-Moody algebra \cite{Kac,FS}.

Now we fix $i\in\{1,\dots,k\}$. From (\ref{3.3}) and (\ref{3.6}) we get
\ber{3.7}
A_{(i)}^{ss'}\in-\Z_+,
\eer
$s,s'\in S_i$, $s \ne s'$.
There are three possibilities for $A_{(i)}$: when a) $\det A_{(i)}>0$,
b) $\det A_{(i)}<0$ and c) $\det A_{(i)}=0$. For $\det A_{(i)}\ne0$ the
corresponding Kac-Moody algebra is simple, since $A_{(i)}$ is
indecomposable \cite{FS}. Now we analyze these three
possibilities.

\subsection{Finite dimensional Lie algebras}

Let $\det A_{(i)}>0$. In this case $A_{(i)}$ is a Cartan matrix of a simple
finite-dimensional Lie algebra and $A_{(i)}^{ss'}\in\{0,-1,-2,-3\}$,
$s\ne s'$. The elements of inverse matrix $A_{(i)}^{-1}$ are positive
(see Ch.7 in \cite{FS}) and hence we get from (\ref{3.4})
\ber{3.8}
\eps_s(U^s,U^s)<0,
\eer
$s\in S_i$.

{\bf Example 1.} Let us consider the Cartan matrix
\ber{3.9}
A_{(i)}=\barr{cc}
2& -q\\
-1& 2
\earr,
\eer
$q=1,2,3$, corresponding to the Lie algebras $A_2=\SL(3)$, $B_2=\SO(5)$
and $G_2$ respectively. Relations (\ref{3.4}) read in this case
\ber{3.10}
\eps_s\nu_s^2(U^s,U^s)(q-4)=4+2q, \ 6
\eer
for $s=1,2$. Here $S_i=\{1,2\}$.

{\bf Example 2.} Let $A_{(i)}$ be $r\times r$ Cartan matrix for the Lie
algebra $A_r= \SL(r+1)$, $r\ge2$. This matrix is described graphically by
the Dynkin diagram pictured on Fig.1.
\begin{center}
\begin{picture}(65,10)
\put(5,5){\line(1,0){34}}
\put(40,5){\makebox(0,0)[lc]{$\dots$}}
\put(47,5){\line(1,0){13}}
\put(5,5){\circle*{1}}
\put(20,5){\circle*{1}}
\put(60,5){\circle*{1}}
\put(5,2){\makebox(0,0)[lc]{1}}
\put(20,2){\makebox(0,0)[lc]{2}}
\put(60,2){\makebox(0,0)[lc]{r}}
\end{picture} \\[5pt]
\small
Fig.1. \it Dynkin diagram for $A_r$ Lie algebra
\end{center}
(For $s\ne s'$, $A_{(i)}^{ss'}=-1$ if nodes $s$ and $s'$ are connected by
a line on the diagram and $A_{(i)}^{ss'}=0$ otherwise). Using the relation
for the inverse matrix $A_{(i)}^{-1}=(A_{ss'}^{(i)})$ (see Ch.7 in
\cite{FS})
\ber{3.11}
A_{ss'}^{(i)}=\frac1{r+1}\min(s,s')[r+1-\max(s,s')]
\eer
we may rewrite (\ref{3.4}) as follows
\ber{3.12}
\eps_s\nu_s^2(U^s,U^s)=s(s-1-r),
\eer
$s\in\{1,\dots,r\}=S_i$.

\subsection{Hyperbolic Kac-Moody algebras}

Now we consider the case $\det A_{(i)}<0$. Among such irreducible
symmetrizable martrices satisfying (\ref{3.7}) there exists a large
subclass of Cartan matrices, corresponding to infinite-dimensional
simple hyperbolic generalized
Kac-Moody (KM) algebras of ranks $r=2,\dots,10$ \cite{Kac,FS}.

{\bf Example 3.} Let
\ber{3.13}
A_{(i)}=\barr{cc}
2& -q_1\\
-q_2& 2
\earr, \quad q_1q_2>4,
\eer
$q_1,q_2\in \N$. This is the Cartan matrix for the hyperbolic KM algebra
$H_{2}(q_1,q_2)$. From (\ref{3.4}) we get
\ber{3.14}
\eps_s\nu_s^2(U^s,U^s)(q_1q_2-4)=2q_s+4,
\eer
$s\in\{1,2\}=S_i$.

{\bf Example 4.} Let $A_{(i)}$ be a Cartan matrix corresponding to
$E_{10}$ hyperbolic KM algebra with the Dynkin diagram pictured
on Fig.2.
\begin{center}
\begin{picture}(90,20)
\put(5,5){\line(1,0){80}}
\put(5,5){\circle*{1}}
\put(15,5){\circle*{1}}
\put(25,5){\circle*{1}}
\put(35,5){\circle*{1}}
\put(45,5){\circle*{1}}
\put(55,5){\circle*{1}}
\put(65,5){\circle*{1}}
\put(75,5){\circle*{1}}
\put(85,5){\circle*{1}}
\put(65,5){\line(0,1){10}}
\put(65,15){\circle*{1}}
\put(5,2){\makebox(0,0)[lc]{1}}
\put(15,2){\makebox(0,0)[lc]{2}}
\put(25,2){\makebox(0,0)[lc]{3}}
\put(35,2){\makebox(0,0)[lc]{4}}
\put(45,2){\makebox(0,0)[lc]{5}}
\put(55,2){\makebox(0,0)[lc]{6}}
\put(65,2){\makebox(0,0)[lc]{7}}
\put(75,2){\makebox(0,0)[lc]{8}}
\put(85,2){\makebox(0,0)[lc]{9}}
\put(68,15){\makebox(0,0)[lc]{10}}
\end{picture} \\[5pt]
\small
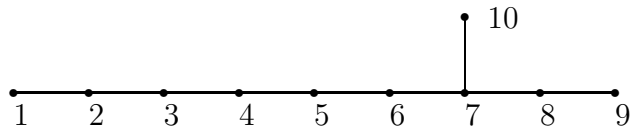
Fig.2. \it Dynkin diagram for $E_{10}$ hyperbolic KM algebra
\end{center}
In this case we get  from (\ref{3.4})  \cite{GrI}
\ber{3.15}
\frac12\eps_s(U^s,U^s)\nu_s^2=30,61,93,126,160,195,231,153,76,115
\eer
for $s=1,2,\dots,10$ respectively.

In both examples for hyperbolic algebras the following
relations are satisfied
\ber{3.16}
\eps_s(U^s,U^s) >0,
\eer
$s\in S_i$. This relation is valid in the general case, since
$(A_{(i)}^{-1})_{ss'}\le0$, $s,s' \in S$, for any hyperbolic algebra
\cite{NikP}.

Hyperbolic KM algebras  appeared in  different areas of
mathematical physics, e.g. in ordinary gravity \cite{FF}
(${\cal F}_3$ hyperbolic algebra),  supergravity:
\cite{J,Miz} ($E_{10}$ hyperbolic algebra),
\cite{Nic} (${\cal F}_3$ hyperbolic algebra),
strings etc
(see also \cite{Nik} and references therein).
In \cite{Nic} it was shown that the  chiral
reduction of a simple ($N=1$) supergravity  from four dimensions
to one dimension gives rise to the hyperbolic algebra of rank
3 (namely ${\cal F}_3$).
In \cite{IKM} an example of cosmological solution
in $D=11$ supergravity describing
three Euclidean $p$-branes (two magnetic
and one electric) with intersection rules
corresponding to the hyperbolic KM algebra ${\cal F}_3$
was constructed.

\subsection{Affine Kac-Moody algebras}

Now we proceed to the degenerate case: $\det A_{(i)}=0$. Here we restrict
ourselves to a subclass of affine KM algebras \cite{Kac,FS}).
Unfortunately, there are no solutions considered
above in the affine case. Indeed, any affine Cartan matrix satisfy the
relations (\ref{3.5}) with $a_s > 0$ called Coxeter labels and, hence, the
solutions are absent in this case.

\section{Solutions with intersecting $p$-branes}
\subsection{The model}

Now we consider a multidimensional gravitational model governed by
the action \cite{IMC}
\ber{4.1}
S=\int d^Dz\sqrt{|g|}\biggl\{R[g]-h_{\alpha\beta}g^{MN}\p_M\varphi^\alpha
\p_N\varphi^\beta-\sum_{a\in\tri}\frac{\theta_a}{n_a!}
\exp[2\lambda_a(\varphi)](F^a)^2\biggr\}
\eer
where $g=g_{MN}dz^M\otimes dz^N$ is the metric,
$\varphi=(\varphi^\alpha)\in\R^l$ is a vector of scalar fields,
$(h_{\alpha\beta})$ is a symmetric
non-degenerate $l\times l$ matrix $(l\in \N)$,
$\theta_a=\pm1$, $F^a=dA^a$ is $n_a$-form ($n_a\ge1$), $\lambda_a$ is a
1-form on $\R^l$: $\lambda_a(\varphi)=\lambda_{\alpha a}\varphi^\alpha$,
$a\in\tri$, $\alpha=1,\dots,l$. Here $\tri$ is some finite set.
In the models with one time all $\theta_a =  1$   when the signature of the
metric  is $(-1,+1, \ldots, +1)$.

We consider the manifold
\ber{4.2}
M=M_0\times M_1\times\dots\times M_n,
\eer
with the metric
\ber{4.3}
g=\e{2\gamma(x)}g^0+\sum_{i=1}^n\e{2\phi^i(x)}g^i
\eer
where $g^0=g_{\mu\nu}^0(x)dx^\mu\otimes dx^\nu$ is a metric on the
manifold $M_0$, and $g^i=g_{m_in_i}^i(y_i)dy_i^{m_i}\otimes dy_i^{n_i}$
is a metric on the manifold $M_i$ satisfying
\ber{4.4}
Ric[g^i]=\xi_ig^i,
\eer
$\xi_i=\const$, $i=1,\dots,n$ ($Ric[g]$ denotes Ricci-tensor,
corresponding to $g$). Thus, all internal spaces $(M_i,g^i)$,
$i=1,\dots,n$, are Einstein ones. Any manifold $M_\nu$ is claimed to
be oriented and connected and $d_\nu\equiv\dim M_\nu$,
$\nu=0,\dots,n$. Let
\ber{4.5}
\tau_i \equiv\sqrt{|g^i(y_i)|}dy_i^1\wedge\dots\wedge dy_i^{d_i}, \quad
\eps(i)\equiv\sign(\det(g_{m_in_i}^i))=\pm1
\eer
denote the volume $d_i$-form and signature parameter respectively,
$i=1,\dots,n$. Let $\Omega=\Omega_n$ be a set of all subsets of
$\{1,\dots,n\}$, $|\Omega|=2^n$. For any $I=\{i_1,\dots,i_k\}\in\Omega$,
$i_1<\dots<i_k$, we denote
\ber{4.6}
\tau(I)\equiv\tau_{i_1}\wedge\dots\wedge\tau_{i_k},
\quad d(I)\equiv\sum_{i\in I}d_i,
\quad \eps(I) \equiv \prod_{i \in I} \eps(i).
\eer
We also put $\tau(\emptyset)= \eps(\emptyset)=
1$ and $d(\emptyset)=0$.

For fields of forms we consider the following composite electromagnetic
ansatz
\ber{4.7}
F^a=\sum_{I\in\Omega_{a,e}}{\cal F}^{(a,e,I)}+
\sum_{J\in\Omega_{a,m}}{\cal F}^{(a,m,J)}
\eer
where
\ber{4.8}
{\cal F}^{(a,e,I)}=d\Phi^{(a,e,I)}\wedge\tau(I), \mm
\label{4.9}
{\cal F}^{(a,m,J)}=\e{-2\lambda_a(\varphi)}*(d\Phi^{(a,m,J)}
\wedge\tau(J))
\eer
are elementary forms of electric and magnetic types respectively,
$a\in\tri$, $I\in\Omega_{a,e}$, $J\in\Omega_{a,m}$ and
$\Omega_{a,e}\subset\Omega$, $\Omega_{a,m}\subset\Omega$. In (\ref{4.9})
$*=*[g]$ is the Hodge operator on $(M,g)$. For scalar functions we put
\ber{4.10}
\varphi^\alpha=\varphi^\alpha(x), \quad
\Phi^s=\Phi^s(x),
\eer
$s\in S$. Here and below
\ber{4.11}
S=S_e \cup S_m, \quad
S_v=\bigcup_{a\in\tri}\{a\}\times\{v\}\times\Omega_{a,v},
\eer
$v=e,m$.

Due to (\ref{4.8}) and (\ref{4.9})
\ber{4.12}
d(I)=n_a-1, \quad d(J)=D-n_a-1,
\eer
for $I\in\Omega_{a,e}$, $J\in\Omega_{a,m}$.

{\bf Remark 1.}
It is more correct to write in (\ref{4.3})
$\hat{g}^{i}$ instead of $g^{i}$, where
$\hat{g}^{i} = p_{i}^{*} g^{i}$ is the
pullback of the metric $g^{i}$  to the manifold  $M$ by the
canonical projection: $p_{i} : M \rightarrow  M_{i}$, $i = 1,
\ldots, n$. Here we omit all ``hats'' (e.g. corresponding to
volume $\tau$-forms) for simplicity.

Let $d_0 \neq 2$ and
\ber{4.13}
\gamma=\gamma_0(\phi) \equiv
\frac1{2-d_0}\sum_{j=1}^n d_j\phi^j,
\eer
i.e. the generalized harmonic gauge is used.

Now we impose restriction on sets $\Omega_{a,v}$.
These restrictions guarantee the block-diagonal structure
of a stress-energy tensor (like for the metric) and the existence of
$\sigma$-model representation \cite{IMC}.

We denote $w_1\equiv\{i|i\in\{1,\dots,n\},\quad d_i=1\}$, and
$n_1=|w_1|$ (i.e. $n_1$ is the number of 1-dimensional spaces among
$M_i$, $i=1,\dots,n$). We also denote
by $I \sqcup  J$ the union of non-intersecting sets $I$ and $J$

{\bf Restriction 1.} For any $a\in\tri$, $v\in\{e,m\}$, there are no
$I,J\in\Omega_{a,v}$ such that
\ber{r.1}
I= \{i\} \sqcup (I \cap J), \qquad J= (I \cap J) \sqcup \{ j \}
\eer
for some $i,j \in w_1$, $i \neq j$.

{\bf Restriction 2} (only for $d_0=1,3$).
For any $a\in\tri$ there are no
$I\in\Omega_{a,m}$, $J\in\Omega_{a,e}$ such that
\ber{r.2}
\bar I=\{i\}\sqcup J
\eer
for $d_0 = 1$ and
\ber{r.3}
J=\{i\}\sqcup \bar I
\eer
for $d_0 = 3$, where $i \in w_1$ and
$\bar I$ is defined as follows
\ber{4.13a}
\bar I\equiv\{1,\ldots,n\}\setminus I.
\eer

Restriction 1  is satisfied for $n_1 \leq 1$ and in
the case when $|\Omega_{a,v}| \leq 1$ for all
$a\in\tri$, $v\in\{e,m\}$ (e.g. in the non-composite case).
For $n_1\ge2$ it forbids certain
pairs of two electric or two magnetic $p$-branes,
corresponding to the same form ($F^a, a \in \tri$).
Restriction 2 is satisfied for $n_1=0$ or when $d_0 \neq 1,3$. For
$n_1\ge1$  and $d_0 = 1,3$ it forbids certain electro-magnetic pairs,
corresponding to the same form.

It was proved in \cite{IMC} that equations of motion for the model
(\ref{4.1}) and the Bianchi identities: $d{\cal F}^s=0$, $s\in S_m$, for
fields from (\ref{4.3})--(\ref{4.13}), when Restrictions 1 and 2 are
imposed, are equivalent to equations of motion for the $\sigma$-model
(\ref{1.1a}) with $(\sigma^A)=(\phi^i,\varphi^\alpha)$, the index set
$S$ from (\ref{4.11}), target space metric
\ber{4.14}
(\hat G_{AB})=\barr{cc}
G_{ij}& 0\\
0& h_{\alpha\beta}
\earr,
\eer
with
\ber{4.15}
G_{ij}= d_i \delta_{ij}+\frac{d_i d_j}{d_0-2},
\eer
the potential
\ber{4.16}
V=-\frac12\sum_{i=1}^n\xi_id_i\e{-2\phi^i+2\gamma_0(\phi)},
\eer
vectors
\ber{4.17}
(U_A^s)=(d_i\delta_{iI_s},-\chi_s\lambda_{\alpha a_s}),
\eer
where $s=(a_s,v_s,I_s)$, $\chi_s=+1,-1$ for $v_s=e,m$ respectively
\ber{4.i}
\delta_{iI}= \sum_{j\in I}\delta_{ij}
\eer
is the indicator of $i$ belonging
to $I$: $\delta_{iI}=1$ for $i\in I$ and $\delta_{iI}=0$ otherwise; and
\ber{4.18}
\eps_s=(-\eps[g])^{(1-\chi_s)/2}\eps(I_s) \theta_{a_s},
\eer
$s\in S$, $\eps[g]\equiv\sign\det(g_{MN})$. More
explicitly (\ref{4.18}) reads
\ber{4.18a}
\eps_s= \eps(I_s) \theta_{a_s}, \quad  v_s= e   \mm
\label{4.18b}
\eps_s= -\eps[g] \eps(I_s) \theta_{a_s} \quad  v_s=m.
\eer

The scalar products (\ref{1.2a}) for vectors $U^s$  were calculated in
\cite{IMC}
\ber{4.19}
(U^s,U^{s'})=d(I_s\cap I_{s'})+\frac{d(I_s)d(I_{s'})}{2-D}+
\chi_s\chi_{s'}\lambda_{\alpha a_s}\lambda_{\beta a_{s'}} h^{\alpha\beta}
\equiv B^{ss'},
\eer
where $(h^{\alpha\beta})=(h_{\alpha\beta})^{-1}$; $s=(a_s,v_s,I_s)$ and
$s'=(a_{s'},v_{s'},I_{s'})$ belong to $S$.

\subsection{Exact solutions with Ricci-flat spaces}

Here we consider the case of Ricci-flat internal spaces, i.e. $\xi_i=0$,
$i=1,\dots,n$, in (\ref{4.4}). The potential (\ref{4.16}) is trivial in
this case and we may apply the results from Sect.2 to our multidimensional
model (\ref{4.1}), when vectors $(U^s,s\in S)$ obey the block-orthogonal
decomposition (\ref{2.3}), (\ref{2.4}) with scalar products defined in
(\ref{4.19}). The solution reads:
\ber{4.20}
g=U\left\{g^0+\sum_{i=1}^n U_ig^i\right\}, \mm
\label{4.21}
U=\left(\prod_{s\in S}H_s^{2d(I_s)\eps_s\nu_s^2}\right)^{1/(2-D)}, \mm
\label{4.22}
U_i=\prod_{s\in S}H_s^{2\eps_s\nu_s^2\delta_{iI_s}}, \mm
\label{4.22a}
Ric[g^0]=Ric[g^1]=\dots=Ric[g^n]=0, \mm
\label{4.23}
\varphi^\alpha=-\sum_{s\in S}\lambda_{a_s}^\alpha\chi_s
\eps_s\nu_s^2\ln H_s, \mm
\label{4.24}
F^a=\sum_{s\in S}{\cal F}^s\delta_{a_s}^a,
\eer
where
\ber{4.25}
{\cal F}^s=\nu_sdH_s^{-1}\wedge\tau(I_s), \mbox{ for } v_s=e, \mm
\label{4.26}
{\cal F}^s=\nu_s(*_0dH_s)\wedge\tau(\bar I_s), \mbox{ for } v_s=m,
\eer
$H_s$ are harmonic functions on $(M_0,g^0)$ coinciding inside blocks
of matrix $(B^{ss'})$ from (\ref{4.19}) ($H_s=H_{s'}$, $s,s'\in S_j$,
$j=1,\dots,k$, see (\ref{2.3}), (\ref{2.4})) and relations
\ber{4.27}
\sum_{s'\in S} B^{ss'} \eps_{s'}\nu_{s'}^2=-1
\eer
on the matrix $(B^{ss'})$,
parameters $\eps_s$ (\ref{4.18}) and  $\nu_s$ are imposed, $s\in S$,
$i=1,\dots,n$; $\alpha=1,\dots,l$. Here $\lambda_a^\alpha=
h^{\alpha\beta}\lambda_{\beta a}$, $*_0=*[g^0]$ is the Hodge operator
on $(M_0,g^0)$ and $\bar I$  is defined in (\ref{4.13a}).

In deriving the solution we used as in \cite{IMC} the relations for
contravariant components of $U^s$-vectors:
\ber{4.29a}
U^{si}=\delta_{iI_s}-\frac{d(I_s)}{D-2}, \quad
U^{s\alpha}=-\chi_s\lambda_{a_s}^\alpha,
\eer
$s=(a_s,v_s,I_s)$.

Thus, we obtained the generalization of the solutions from \cite{IMC} to the
block-orthogonal case (here we eliminate the misprint with sign in eq.
(5.19) in \cite{IMC}).

{\bf Remark 2}. The solution is also valid for $d_0=2$, if Restriction 2
is replaced by Restriction $2^{*}$. It may be proved using a more general
form of the sigma-model representation (see Remark 2 in \cite{IMC}).

{\bf Restriction 2${}^*$} (for $d_0=2$). For any  $a \in \tri$  there are
no $I\in\Omega_{a,m}$, $J\in\Omega_{a,e}$ such that $\bar I =  J$ and for
$n_1 \geq 2$, $i,j\in w_1$, $i \neq j$, there are no $I\in\Omega_{a,m}$,
$J\in\Omega_{a,e}$ such that $i\in I$, $j\in \bar J$, $I\setminus\{i\}=
\bar J \setminus \{j\}$.

\subsection{Behaviour of the Kretschmann scalar and horizon}

Let $M_0=\R^{d_0}$, $d_0>2$ and $g^0=\delta_{\mu\nu}dx^\mu\otimes dx^\nu$.
For
\ber{4.28}
H_s=1+\sum_{b\in X_s}\frac{q_{sb}}{|x-b|^{d_0-2}},
\eer
where $X_s$ is finite non-empty subset $X_s\subset M_0$, $s\in S$, all
$q_{sb}>0$, and $X_s=X_{s'}$, $q_{sb} = q_{s'b}$ for $b \in X_s=X_{s'}$,
$s, s' \in S_j$, $j =1, \ldots, k$.  The harmonic functions (\ref{4.28}) are
defined in domain $M_0\setminus X$, $X=\bigcup_{s\in S}X_s$,
and  generate the solutions (\ref{4.20})--(\ref{4.27}).

Denote $S(b)\equiv\{s\in S|\quad b \in X_s\}$. We also put  $M_i=\R^{d_i}$,
$R[g^i]= {\cal K}[g^i]=0$, $i=1,\dots,n$,
where
\ber{4.k}
{\cal K}[g] \equiv R_{MNPQ}[g]R^{MNPQ}[g]
\eer
is the Kretschmann scalar (or Riemann tensor squared). Then for the metric
(\ref{4.20}) we obtain
\ber{4.29}
{\cal K}[g] = \frac{C' + o(1)}{U^2|x-b|^4}
= [C + o(1)]|x-b|^{4(d_0-2)\eta(b)}
\eer
for $x\to b \in X$, where
\ber{4.30}
\eta(b)\equiv\sum_{s\in S(b)}(-\eps_s)\nu_s^2 \frac{d(I_s)}{D-2}-
\frac{1}{d_0-2},
\eer
and $C = C(b) \geq 0$ ($C = {\rm const}$). The relation for
$C = C(b)$ is given in Appendix.
Relation (\ref{4.29}) may be obtained using the relation for
${\cal K}[g]$ in Appendix of Ref. \cite{IMA}.
In what follows we consider
non-exceptional $b \in X$ defined by relations $C = C(b) > 0$.

{\bf Remark 3.}
It follows from the Appendix  that an exceptional point $b \in X$,
defined by relation $C = C(b) = 0$, appears iff (if and only if)
\ber{4.30a}
U(x) \sim c |x-b|^{ -2 \alpha}, \quad   U(x) U_i(x)  \sim c_i,
\eer
for $x\to b$, where  $\alpha = 0,2$ and $c, c_i \neq 0$
are constants, $i=1,\dots, n$.

Due to (\ref{4.29}) the metric (\ref{4.20}) has no
curvature singularity when $x\to b \in X$, $C(b) > 0$, iff
\ber{4.31}
\eta(b)\ge 0.
\eer

>From (\ref{4.30}) we see that the metric (\ref{4.20})
is regular at a ``point'' $b\in X$
for $\eps_s=-1$ and large enough values of $\nu_s^2$, $s\in S(b)$. For
$\eps_s=+1$, $s\in S(b)$, we have a curvature singularity at
non-exceptional point $b\in X$.

Now we consider a special case: $d_1=1$,
$g^1=-dt\otimes dt$.
In this case we have a horizon, when $x\to b\in X$, iff
\ber{4.32}
\xi(b)\equiv \sum_{s\in S(b)}(-\eps_s) \nu_s^2 \delta_{1I_s}
-\frac1{d_0-2}\ge0.
\eer
This relation follows from the requirement of infinite time
propagation of light to $b \in X$.
If $\eps_s= -1$, $1 \in I_s$ for all $s\in S(b)$, we get
\ber{4.33}
\eta(b)<\xi(b),
\eer
$b\in X$. This follows from the inequalities $d(I_s)<D-2$ ($d_0>2$).

We note that $g_{tt} \to 0$
for $x \to b \in X$, if (\ref{4.33}) is satisfied.
This follows from the relation
\ber{4.33a}
g_{tt} \sim {\rm const} |x-b|^{2(d_0-2)(\xi(b)- \eta(b))},
\eer
$x \to b$.

{\bf Remark 4.}
Due to relations (\ref{4.30a}) and (\ref{4.33a})
the point $b \in X$ is non-exceptional if
$g^1=-dt\otimes dt$ and $1\in I_s$, $\eps_s= -1$ for
all $s \in S(b)$.

Thus, for the metric (\ref{4.20}) with $H_s$ from (\ref{4.28})
there are two dimensionless indicators at the
non-exceptional point $b\in X$:
a) horizon indicator $\xi(b)$ (corresponding to time $t$)
and b) curvature singularity indicator
$\eta(b)$.  These indicators  define (for our assumptions)
the existence of a horizon
and the singularity of the Kretschmann scalar (when $(M_i,g^i)$ are flat,
$i=1, \ldots,n$) at non-exceptional $b \in X$.

\subsection{Generalized MP solutions}

Here we consider special black hole solutions
for the model  (\ref{4.1})   with  all $\theta_a =  1$, $a \in \tri$,
when the signature of the
metric  $g$ is $(-1,+1, \ldots, +1)$.
>From the results of the previous subsections we get the
following  solution (with all $\eps_s = \eps(I_s)= -1$)
\ber{4.34}
g= \Bigl(\prod_{s \in S} H_s^{2d(I_s)\nu_s^2} \Bigr)^{1/(D-2)}
\Bigl\{ \sum_{\mu=1}^{d_0} dx^{\mu} \otimes dx^{\mu}  \nn
- \Bigl(\prod_{s \in S} H_s^{-2 \nu_s^2 }\Bigr) dt \otimes dt
+ \sum_{i=2}^{n}  \Bigl( \prod_{s\in S}
H_s^{-2\nu_s^2\delta_{iI_s}}\Bigr) g^i \Bigr\}, \\
\label{4.35}
\varphi^\alpha= \sum_{s\in S}\lambda_{a_s}^\alpha\chi_s
\nu_s^2\ln H_s, \\
\label{4.36}
F^a=
\sum_{s\in S_e} \delta_{a_s}^a \nu_s dH_s^{-1} \wedge \tau(I_s)
+ \sum_{s\in S_m} \delta_{a_s}^a \nu_s (*_0 dH_s) \wedge \tau(\bar{I}_s),
\eer
where $(M_i,g^i)$  are flat Euclidean spaces, $i = 2, \ldots, n$,
\ber{4.37}
1 \in I_s,
\eer
$s \in S$, i.e. all branes have a common time submanifold $M_1 = \R$,
the harmonic functions $H_s$ (\ref{4.28}) are coinciding inside blocks,
i.e. $H_s=H_{s'}$, $s,s'\in S_j$, $j=1,\dots,k$,
parameters $\nu_s$ satisfy the relations
\ber{4.27a}
\sum_{s'\in S} B^{ss'} \nu_{s'}^2=1
\eer
with
the  matrix $(B^{ss'})$ from (\ref{4.19}),
$*_0=*[g^0]$ is the Hodge operator on $(M_0 = \R^{d_0},
g^0 = \sum_{\mu=1}^{d_0} dx^{\mu} \otimes dx^{\mu})$,
$\bar I$  is defined in (\ref{4.13a}), $\theta_a = 1$
for all $a \in \tri$, and
\ber{4.38}
\eta(b) = \sum_{s\in S(b)}\nu_s^2\frac{d(I_s)}{D-2}-
\frac1{d_0-2} \geq 0,
\eer
$b \in X$. This solution describes a set of extreme
$p$-brane black holes with horizons at $b \in X$.
The Riemann tensor squared has a finite limit
at any $b \in X$.

Calculation of the Hawking temperature
corresponding to $b\in X$ using standard formula (see, for
example, \cite{W,BIM})
gives us
\ber{4.39}
T_H(b)=0,
\eer
for any $b\in X$ satisfying $\xi(b) > 0$.

{\bf MP solution.} The standard 4-dimensional
Majumdar-Papapetrou solution \cite{MP} in our notations reads
\ber{4.40}
g=H^2g^0-H^{-2}dt\otimes dt, \mm
\label{4.41}
F=\nu dH^{-1} \wedge dt,
\eer
where $\nu^2 = 2$, $g^0 = \sum_{i=1}^{3} dx^i \otimes dx^i$
and $H$ is a harmonic function. We have one electric $0$-brane (point)
``attached'' to the time manifold; $d(I_s) =1$, $\eps_s= -1$ and
$(U^s,U^s) = 1/2$. In this case
(e.g. for extremal Reissner-Nordstr\"om black hole) we get
\ber{4.41a}
\eta(b)=0, \qquad \xi(b)=1,
\eer
and $T_H(b)=0$, $b\in X$.

{\bf D = 11 supergravity.} Let us consider the
action (\ref{4.1})
without scalar fields and with the only one form $F^4$ of
rank $4$. This action coincides with  the truncated (i.e.
without Chern-Simons term) bosonic part of $D=11$
supergravity action. In this case
there are a lot of  $p$-brane
MP-type (multicenter extremal black hole) solutions
with orthogonal intersection rules \cite{DS}-\cite{GKT}, e.g.
(i) solution with one electric $2$-brane ($d(I_s)=3$) and
$d_0 =8$;
(ii) solution with one magnetic $5$-brane ($d(I_s)=6$) and
$d_0 =5$;
(iii) solution with  one electric $2$-brane and one
 magnetic $5$-brane ($d(I_{s_1}\cap I_{s_2})=2$) and
 $d_0 =4$;
(iv) solution with  two electric $2$-branes
    ($d(I_{s_1}\cap I_{s_2})=1$) and $d_0 =5$;
(v) solution with  two magnetic $5$-branes
    ($d(I_{s_1}\cap I_{s_2})= 4$) and $d_0 =3$.
In the examples  (iii)-(v) the harmonic functions
$H_{s_1}$ and $H_{s_2}$ from (\ref{4.28}) should
have the coinciding sets of poles, i.e. $X_{s_1} = X_{s_2}$,
to maintain the relation (\ref{4.38}). The addition of the
Chern-Simons term does not destroy these solutions.
In all these examples $\eta(b)=0$, $b \in X_{s}$ and
$\nu_s^2 = 1/2$, $s \in S$.

\subsection{Non-extremal $p$-brane black hole solutions}

There exists a non-extremal generalization
of the solution from the previous subsection,
when all the poles in all harmonic functions are coinciding,
say $b =0$, $b \in X$.

The metric and scalar fields for this solution read
\ber{4.42}
g=
\Bigl(\prod_{s \in S} H_s^{2 d(I_s)\nu_s^2/(D-2)} \Bigr)
\biggl\{ \frac{dr \otimes dr}{1 - 2\mu / r^{\bar d}}
+ r^2  d \Omega^2_{d_0 -1} \nn    \nn
- \Bigl(\prod_{s \in S} H_s^{-2  \nu_s^2} \Bigr)
 \left(1 - \frac{2\mu}{r^{\bar d }} \right)  dt \otimes dt
+ \sum_{i = 2}^{n} \Bigl(\prod_{s\in S}
  H_s^{-2 \delta_{iI_s} \nu_s^2} \Bigr) g^i  \biggr\}, \\
\label{4.43}
\varphi^\alpha=
\sum_{s\in S} \nu_s^2 \chi_s \lambda_{a_s}^\alpha
\ln H_s,
\eer
where
\beq{4.44}
H_s = 1 + \frac{q_s}{r^{\bar{d}}},
\eeq
$s \in S$, $\bar{d} = d_0 -2$, $\mu > 0$, parameters
$\nu_s$ satisfy the relations (\ref{4.27a})
with the matrix $(B^{ss'})$ from (\ref{4.19}),
and parameters $q_s > 0$ coincide inside blocks
i.e. $q_s=q_{s'}$, $s,s'\in S_j$, $j=1,\dots,k$.
Here $(M_i,g^i)$  are Ricci-flat Euclidean spaces, $i = 2, \ldots, n$,
and $d \Omega^2_{d_0 -1}$ is canonical metric
on the $(d_0 -1)$-dimensional sphere $S^{d_0 -1}$.
The fields of forms are given by
(\ref{4.24})-(\ref{4.26})
with
\ber{4.45}
\Phi^s = \frac{\nu_s}{H_s^{'} },  \mm
\label{4.46}
H_s^{'}= \Bigl(1 - \frac{p_s}{H_s r^{\bar{d}}} \Bigr)^{-1} =
1 + \frac{ p_s}{ r^{\bar{d}} +  q_s -  p_s}, \mm
\label{4.47}
p_s = \sqrt{q_s (q_s + 2\mu)},
\eer
$s \in S$.

The solution  describes non-extremal charged
intersecting $p$-brane black hole
with the horizon at  $r^{\bar{d}} = 2\mu$.
In the limit $\mu \to + 0$
this solution coincides with the
1-center  extremal black hole solution
>from the previous subsection.
Here we generalize the solution from
\cite{Br}, where all $\nu_s^2$ are coinciding inside blocks.
In our case the only restriction on $\nu_s$ is the relation
(\ref{4.27a}).
For orthogonal intersections it agrees with the solutions
>from \cite{CT}-\cite{O} ($d_1 = \ldots = d_n =1$) and \cite{BIM,IMJ}.

The Hawking temperature corresponding to
this non-extremal solution is
\beq{4.48}
T_H(0,\mu) =
\frac{\bar{d}}{4 \pi (2 \mu)^{1/\bar{d}}}
\prod_{s \in S}
\left(\frac{2 \mu}{2 \mu +  q_s}\right)^{\nu_s^2}.
\eeq
For $\mu \to + 0$ we get (in agreement with (\ref{4.39}))
$T_H(0,\mu) \to 0$ for the extremal
black hole configurations satisfying
\beq{4.49}
\xi(0) = \sum_{s\in S}\nu_s^2-
\bar{d}^{-1} > 0.
\eeq

\section{Intersection rules and some examples}
\subsection{Intersection rules}

>From orthogonality relation (\ref{2.4}) and (\ref{4.19}) we get
\ber{5.1}
d(I_s \cap I_{s'})=\tri(s,s')
\eer
where $s\in S_i$, $s'\in S_j$, $i\ne j$ and
\ber{5.2}
\tri(s,s')\equiv\frac{d(I_s)d(I_{s'})}{D-2}-
\chi_s\chi_{s'}\lambda_{a_s}\cdot\lambda_{a_{s'}}.
\eer
Here $\lambda\cdot\lambda'\equiv h^{\alpha\beta}\lambda_\alpha\lambda'_\beta$.
Let
\ber{5.3}
N(a,b)\equiv\frac{(n_a-1)(n_b-1)}{D-2}-\lambda_a\cdot\lambda_b,
\eer
$a,b\in\tri$. The matrix (\ref{5.3}) is called
the fundamental matrix of the
model (\ref{4.1}) \cite{IMJ}. For $s_1,s_2\in S$, $s_1\ne s_2$, the
$\tri$-symbol (\ref{5.2}) may be expressed by means of the
fundamental matrix \cite{IMJ}
\ber{5.4}
\tri(s_1,s_2)=\bar D\bar\chi_{s_1}\bar\chi_{s_2}+
\bar{n}_{a_{s_1}}\chi_{s_1}\bar\chi_{s_2}+
\bar{n}_{a_{s_2}}\chi_{s_2}\bar\chi_{s_1}+
N(a_{s_1},a_{s_2})\chi_{s_1}\chi_{s_2},
\eer
where $\bar D=D-2$, $\bar n_a=n_a-1$, $\bar\chi_s=\frac12(1-\chi_s)$.
More explicitly (\ref{5.4}) reads
\ber{5.4a}
\Delta(s_1,s_2)
=N(a_{s_1},a_{s_2}), \quad v_{s_1}=v_{s_2}=e; \mm
\label{5.4b}
\Delta(s_1,s_2) =
\bar{n}_{a_{s_1}}-N(a_{s_1},a_{s_2}), \quad v_{s_1}=e, \quad v_{s_2}=m; \mm
\label{5.4c}
\Delta(s_1,s_2) =
\bar{D}-\bar{n}_{a_{s_1}}-\bar{n}_{a_{s_2}}+N(a_{s_1},a_{s_2}),
\quad v_{s_1}=v_{s_2}=m.
\eer

This follows from the relations
\ber{5.5}
d(I_s)=\bar D\bar\chi_s+\bar n_{a_s}\chi_s,
\eer
equivalent to (\ref{4.12}). Let
\ber{5.6}
K(a)\equiv n_a-1-N(a,a)=\frac{(n_a-1)(D-n_a-1)}{D-2}+
\lambda_a\cdot\lambda_a,
\eer
$a\in\tri$.

The parameters (\ref{5.6}) play a rather important role in supergravity
theories, since they are preserved under Kaluza-Klein reduction \cite{St}
and define the norms of $U^s$ vectors:
\ber{5.7}
(U^s,U^s)=K(a_s),
\eer
$s\in S$.

Here we put $K(a)\ne0$, $a\in\tri$. Then, we obtain the general intersection
rule formulas
\ber{5.8}
d(I_{s_1}\cap I_{s_2})=\tri(s_1,s_2)+\frac12K(a_{s_2})A^{s_1s_2}
\eer
$s_1\ne s_2$, where $(A^{s_1s_2})$ is the quasi-Cartan matrix (\ref{3.2})
(see also (6.32) from \cite{IMJ}).

\subsection{$B_D$-models and examples of solutions}

Now we consider some examples of solutions from Sect. 4. These examples
will be demonstrated for the so-called $B_D$-models. Action
of the $B_D$-model reads \cite{IMJ}
\ber{5.9}
S_D=\int d^Dz\sqrt{|g|}\biggl\{R[g]+g^{MN}\p_M\vec\varphi\p_N\vec\varphi-
\sum_{a=4}^{D-7}\frac1{a!}\exp[2\vec\lambda_a\vec\varphi](F^a)^2\biggr\},
\eer
where $\vec\varphi=(\varphi^1,\dots,\varphi^l)\in\R^l$, $\vec\lambda_a=
(\lambda_{a1},\dots,\lambda_{al})\in\R^l$, $l=D-11$, $\rank F^a=a$,
$a=4,\dots,D-7$. Here vectors $\vec\lambda_a$ satisfy the relations
\ber{5.10}
\vec\lambda_a\vec\lambda_b=N(a,b)-\frac{(a-1)(b-1)}{D-2}, \mm
\label{5.11}
N(a,b)=\min(a,b)-3,
\eer
$a,b=4,\dots,D-7$.

The vectors $\vec\lambda_a$ are linearly dependent
\ber{5.12}
\vec\lambda_{D-7}=-2\vec\lambda_4.
\eer
For $D>11$ vectors $\vec\lambda_4,\dots,\vec\lambda_{D-8}$ are linearly
independent.

The model (\ref{5.9}) contains $l$ scalar fields with a negative kinetic
term (i.e. $h_{\alpha\beta}=-\delta_{\alpha\beta}$ in (\ref{4.1}))
coupled to $(l+1)$ forms. For $D=11$ ($l=0$) the model (\ref{5.9})
coincides with a truncated (without Chern-Simons term) bosonic sector
of $D=11$ supergravity. For $D=12$ $(l=1)$ (\ref{5.9}) coincides with
truncated $D=12$ model from \cite{KKP} (see also \cite{IMC}).

The matrix (\ref{5.11}) is the fundamental matrix of the $B_D$-model.

For $p$-brane worldsheets we have the following dimensions (see
\cite{IMJ})
\ber{5.13}
d(I)=3,\dots,D-8, \quad I\in\Omega_{a,e}, \mm
\label{5.14}
d(I)=D-5,\dots,6, \quad I\in\Omega_{a,m}.
\eer
Thus, there are $(l+1)$ electric and $(l+1)$ magnetic $p$-branes, $p=d(I)-1$.
For $B_D$-model all $K(a)=2$.

Now we are interested in the one-block solutions, where
$A=(A^{s_1s_2})$ in (\ref{3.2}) is an irreducible
Cartan matrix either of a simple
finite dimensional Lie algebra or of a simple hyperbolic KM algebra.

Since $K(a)=2$, we rewrite (\ref{5.8}) as the following:
\ber{5.13a}
d(I_{s_1}\cap I_{s_2})=\tri(s_1,s_2)+A^{s_1s_2},
\eer
$s_1\ne s_2$, and get $A^{s_1s_2}=A^{s_2s_1}$, i.e. the Cartan matrix is
symmetric.

\subsubsection{Finite dimensional Lie algebras}

Here we put all $\theta_a =  1$, $a \in \tri$.
>From $\eps_s=-1$ we get $\eps(I_s)=-1$ for  the electric "brane" and
$\eps(I_s)= \eps[g]$ for the magnetic one. In a finite dimensional case we
are led to the so-called simply laced or $A-D-E$ Lie algebras. The
intersection rules are totally defined by the corresponding Dynkin diagram:
$d(I_{s_1}\cap I_{s_2})= \tri(s_1,s_2)-1$, when the vertices corresponding
to $s_1$ and $s_2$ are connected by a line and $d(I_{s_1}\cap
I_{s_2})=\tri(s_1,s_2)$ otherwise (since in $A-D-E$ case $A^{s_1s_2}=0, -1$,
$s_1 \neq s_2$).

{\bf Example for $A_2$.} Let us consider $B_D$-model, $D\ge11$,
$a\in\{4,\dots,D-7\}$, $g^3=-dt\otimes dt$, $d_1=a-2$, $d_2=D-2-a$,
$d_0=3$ and $g^0,g^1,g^2$ are Ricci-flat. The
$A_2$-solution describing a dyon configuration with electric $d_1$-brane
and magnetic $d_2$-brane, corresponding to $F^a$-form and
intersecting in 1-dimensional time manifold  reads:
\ber{5.14a}
g=H^2g^0-H^{-2}dt\otimes dt+g^1+g^2, \mm
\label{5.15}
F^a=\nu_1dH^{-1}\wedge dt\wedge\tau_1+\nu_2(*_0dH)\wedge\tau_1,
\eer
$\vec\varphi=0$, $H$ is the harmonic function on $(M_0,g^0)$
and $\nu_1^2 = \nu_2^2 =1$. For $D=11$
we have $a=4$ and $d_1=2$, $d_2=5$. For $D=12$ we have two possibilities:
a) $a=4$, $d_1=2$, $d_2=6$; b) $a=5$, $d_1=3$, $d_2=5$. The signature
restrictions on $g^1$ and $g^2$ are the following: $\eps(1)=+1$,
$\eps(2) = -\eps[g]$. They are satisfied when $g^0$ and $g^1$
are Euclidean metrics.
The 4-dimensional section of (\ref{5.14a}) coincides for the flat Euclidean
$g^0$ with the MP solution  (\ref{4.40}), (\ref{4.41}). Here the
"indicators" of the solution coincide with MP ones (\ref{4.41a}).

Now, we list $A_2$ intersection rules, for $D=11,12$. Here
\ber{5.15a}
A=B=\barr{cc}
2& -1\\
-1& 2
\earr.
\eer
${\bf D=11}$ :
\ber{5.17a}
3\cap3=0, \quad 3\cap6=1, \quad 6\cap6=3.
\eer
${\bf D=12}$ :
\ber{5.17b}
3\cap3=3\cap4=0, \quad 3\cap6=3\cap7=4\cap4=4\cap6=1, \nn
4\cap7=2, \quad 6\cap6 = 6\cap7= 3, \quad 7\cap7=4.
\eer
Here and in what follows we denote $n_1\cap n_2=n$ $\Leftrightarrow$
($d(I_1)=n_1$, $d(I_2)=n_2$, $d(I_1\cap I_2)=n$).

{\bf Remark 5.}  The appearance of Lie algebras
it is not accident here. It was shown in \cite{IMJ}
that if $p$-branes have intersection
rules   (\ref{5.8}) governed by a Cartan matrix of some
(semisimple) Lie algebra, then equations of motion
for the problems of cosmology and spherical symmetry may
be reduced to integrable Euclidean Toda lattice equations
corresponding to this Lie algebra. (For certain examples of solutions
see, for example, \cite{LPX,LMPX,LMMP,GMel}.)
Thus, for intersections related to Lie algebras
we may find general spherically symmetric solutions
which contain one-center MP-type solutions
(e.g. extremal black hole configurations)
as  special case.

\subsubsection{Hyperbolic algebras}

In hyperbolic case all $\eps_s=+1$, $s\in S_1 =S$ and, hence, corresponding
solutions with $H_s$ from (\ref{4.28}) are singular at $b \in X$
(in this case $b$ is non-exceptional).

{\bf Example with $H_2(q,q)$.} For the Cartan matrix (\ref{3.13}) with
$q_1=q_2=q \ge 3$, we obtain
\ber{5.16}
\nu_s^2=(q-2)^{-1}=1,\frac12,\frac13,\dots
\eer
for $q=3,4,5,\dots$; $s=1,2$. An example of the solution for $d_0=3$ with
two electric $p$-branes, $p=d_1,d_2$,
corresponding to $F^a$ and $F^b$ fields and intersecting in time manifold,
is the following:
\ber{5.17}
g=H^{-2/(q-2)}g^0-H^{2/(q-2)}dt\otimes dt+g^1+g^2, \mm
\label{5.18}
F=\nu_{1}dH^{-1}\wedge dt\wedge\tau_1+
\nu_{2}dH^{-1}\wedge dt\wedge\tau_2, \mm
\label{5.19}
\vec\varphi=-(\vec\lambda_a+\vec\lambda_b)(q -2)^{-1}\ln H
\eer
where $d_1=a - 2$, $a=q+4$, $b\ge a$, $d_2=b-2$, $d_0 = 3$, $D=a+b$. Here
$F=F^a+F^b$ for $a<b$ and $F=F^a$ for $a=b$. The signature
restrictions are : $\eps(1)= \eps(2) = -1$. Thus, the space-time
$(M,g)$ should contain at least three time directions.
The minimal $D$ is 14. For $D=14$ we get $a=b=7$, $d_1=d_2=6$,
$q=3$. In this case $6\cap6=1$.

\subsubsection{Affine (forbidden) case}

Affine Cartan matrices do not arise in our solutions. This means
that some configurations are forbidden. Let us consider $A_1^{(1)}$
affine KM algebra with the Cartan matrix
\ber{5.20}
A=\barr{cc}
2& -2\\
-2& 2
\earr.
\eer
For $D=11$ the intersections: $3\cap6=0$, $6\cap6=2$, corresponding
to the $A$-matrix (\ref{5.20}), are forbidden. For $D=12$
we get forbidden intersections: $3\cap6=3\cap7=4\cap4=4\cap6=0$,
$4\cap7=1$, $6\cap6=6\cap7=2$, $7\cap7=3$.

{\bf Remark 6.} Recently some new solutions in the affine
case were obtained \cite{ETT}. (These solutions contain as a
special case a solution in $D =11$ supergravity from \cite{GKT}
with $6 \cap 6=2$.) The solutions from \cite{ETT}
use some modified  ansatz for fields of forms
and  do not belong to our scheme. This indicates that
the sigma-model  considered in this paper is not the
most general one describing possible $p$-brane configurations.
There are at least  three possible ways of
generalization:
(i)  ``non-block-diagonal'' metric instead of
the ``block-diagonal'' metric (\ref{4.3}) may be considered
(the first step in this direction was done in \cite{GR});
(ii)  a bimetric sigma-model defined on
the product of  two base spaces  ($M_{01},g^{01}$) and ($M_{02},g^{02}$)
instead of ($M_{0},g^{0}$) with two  sorts
of $p$-branes governed by functions on $M_{01}$ and $M_{02}$
respectively may be constructed ;
(iii) a sigma-model describing the
action (\ref{4.1}) with Chern-Simons terms added
may be also considered.
Solution from \cite{ETT} seems to belong to  the case (ii).
We note also that our solutions in the special case of
$D= 10,11$ supergravities are
different from non-marginal bound state
solutions (see, for example, \cite{O,ILPT,RT}
and references therein) although the rules for binary intersections
may look similar. These solutions probably
need some extensions of the sigma-model along the lines (i) and (iii).

\subsubsection{Other possibilities}

We note that it is not obvious for quasi-Cartan matrix $A$
to be a Cartan one. Let us consider the example with
\ber{5.21}
A=\barr{cc}
2& 1\\
1& 2
\earr.
\eer
For $D=11$ we obtain (non-Lie) dyon with
\ber{5.22}
3\cap6=3
\eer
and $\eps_s=-1$, $\nu_s^2=1/3$, $s =1,2$ (see also \cite{Br}). Two other
intersections: $3\cap3=2$, $6\cap6=5$, corresponding to $A$
are forbidden by Restriction 1 (this restriction
follows from the block-diagonal form of metric $g$ and may be
weakened for non-block-diagonal ansatz from \cite{GR}).
For $d_0= 5$ the solution
with $d_1=d_2=3$ reads
\ber{5.23}
g=H^{2/3}g^0+H^{-2/3}g^1+g^2, \nn
F^4 = \nu_1dH^{-1}\wedge\tau_1+\nu_2*_0dH
\eer
with $H$ from (\ref{4.28}) and indicators $\xi(b)=1/3$, $\eta(b)=0$,
$b\in X$, i.e. any point $a\in X$ corresponds to the horizon without
a curvature singularity. Here we assume that $M_1 = \R \times M_1^{'}$,
where $\R$ is the time manifold.

\section{Conclusions}

Thus, here, like in \cite{IM4}-\cite{IMR}, we considered the
$\sigma$-model of a   $p$-brane origin. We obtained new
block-orthogonal exact solutions (\ref{2.8})--(\ref{2.12})
governed by a set of harmonic functions. These solutions crucially
depend on the matrix of scalar products $(U^s,U^{s'})$, $s,s'\in S$, and
parameters $\eps_s=\pm1$, $s\in S$ (see (\ref{2.10})).
In Sect. 3 we analyzed
three possibilities, when quasi-Cartan matrix (\ref{3.2}) coincides with
the Cartan matrix of a (simple): a) finite-dimensional Lie algebra;
b) hyperbolic  KM algebra; c) affine KM algebra. It was shown that
the last possibility does
not appear in our solutions, and all $\eps_s=-1$ in the case a) and all
$\eps_s=+1$ in the case b). In Sect. 4 we applied
this method for obtaining new
MP type solutions in multidimensional gravity with fields
of forms and scalar fields (see subsect. 4.2 and 4.4).
In subsect. 4.3 for $d_0>2$ and
harmonic functions from (\ref{4.28}) the  indicators $\eta(b)$ and
$\xi(b)$ were introduced. These indicators describe
under certain assumptions the existence of a
curvature singularity and a horizon for $x\to b$. In Sect. 5 intersection
rules for fixed $A$-matrix are written  (see (\ref{5.8}))  and some examples
of solutions and intersection rules for a chain of $B_D$-models,
$D=11,12,\dots$, were suggested. Among the examples there are
$A_2$, hyperbolic, affine (forbidden) intersections and ``non-Cartan'' ones.
We note that the solutions
obtained here may be also generalized to the case when some non-Ricci-flat
internal spaces are added  to $M$ as it was done in \cite{IMC}.

\section{Appendix}

Here we present the relation for the parameter
$C = C(b)$, $b \in X$, from (\ref{4.29})
\ber{a.1}
C = C_0 + C_1 + C_2, \mm
\label{a.2}
C_0 = 2 (d_0-1)(d_0 -2) \alpha^2 (\alpha -2)^2, \mm
\label{a.3}
C_1 = 4 [(d_0 - 1) \alpha^2  + (\alpha - 1)^2]
\sum_{i=1}^{n} d_i \alpha_i^2, \mm
\label{a.4}
C_2 = 2 (\sum_{i=1}^{n} d_i \alpha_i^2)^2 - 2
\sum_{i=1}^{n} d_i \alpha_i^4,
\eer
where
\ber{a.5}
\alpha = \alpha(b) \equiv
(d_0-2) \sum_{s\in S(b)}(-\eps_s)\nu_s^2 \frac{d(I_s)}{D-2}
= (d_0 -2) \eta(b) + 1, \mm
\label{a.6}
\alpha_i = \alpha_i(b) \equiv
(d_0-2) \sum_{s\in S(b)}(-\eps_s)\nu_s^2
\left[\delta_{iI_s} - \frac{d(I_s)}{D-2} \right],
\eer
$i=1, \dots,n$.

It follows from definitions  (\ref{a.1})-(\ref{a.4})
that $C \geq 0$ and
\ber{a.7}
C =0 \Leftrightarrow (\alpha = 0, 2, \quad \alpha_i =0, \ i=1, \dots,n).
\eer

Parameter $C$ appears in the Kretschmann scalar
(\ref{4.29}) for the ``1-pole'' metric
\ber{a.8}
g_*= r^{-2 \alpha}[dr \otimes dr + r^2  d \Omega^2_{d_0 -1} ]
+ \sum_{i = 1}^{n} r^{2\alpha_i} g^i,
\eer
with $R[g^i]= {\cal K}[g^i]=0$, $i=1,\dots,n$.
Using formulas from Appendix of \cite{IMA},
we obtain
\ber{a.9}
{\cal K}[g_*] = C r^{-4 +4 \alpha}.
\eer

\begin{center}
{\bf Acknowledgments}
\end{center}

This work was supported in part
by the DFG grant
436 RUS 113/236/O(R),
by the Russian Ministry of
Science and Technology and  Russian Foundation for Basic Research.
The authors  are grateful to M.A. Grebeniuk, K.A. Bronnikov,
S.-W. Kim  for useful discussions and V.V. Nikulin for
valuable  information.
One of us (V.D.I) also thanks  S.-W. Kim for a kind
hospitality during his stay in Ewha Womans University (Seoul, S. Korea).

\end{document}